\documentclass[twocolumn,prl,showpacs]{revtex4}

\usepackage{graphicx}
\usepackage{amssymb}
\usepackage{amsfonts}

\DeclareGraphicsExtensions{.eps}

\begin{document}

\title{Reversible Quantum Interface for Tunable Single-sideband Modulation}

\author{J. Cviklinski, J. Ortalo, J. Laurat, A. Bramati, M. Pinard, E. Giacobino$^\ast$}

\address{Laboratoire Kastler Brossel, Universit\'{e} Pierre et Marie Curie, Ecole Normale Supérieure, CNRS,
4 place Jussieu, F75252 Paris Cedex 05, France.}

\begin{abstract}
Using Electromagnetically Induced Transparency (EIT) in a Cesium
vapor, we demonstrate experimentally that the quantum state of a
light beam can be mapped into the long lived Zeeman coherences of
an atomic ground state. Two non-commuting variables carried by
light are simultaneously stored and subsequentely read-out, with
no noise added. We compare the case where a tunable single
sideband is stored independently of the other one to the case
where the two symmetrical sidebands are stored using the same EIT
transparency  window.

\end{abstract}


\maketitle

Developing memory registers for quantum signals carried by light is
an essential milestone for quantum information processing
\cite{EPJD}. Atomic ensembles are good candidates for such registers
since quantum states of light can be stored in long-lived atomic
spin states and retrieved on demand. Several protocols have been
proposed for such memories, based on three-level systems interacting
with two light fields, in a Raman-type configuration
\cite{Polzik1,DLCZ,Dantan1} and in an electromagnetically induced
transparency (EIT) configuration
\cite{Lukin1,Lukin2,Dantan2,Dantan3}, or based on a quantum non
demolition (QND) interaction \cite{Polzik2}. Substantial advances
have been made in the single photon regime, including the storage of
single photons \cite{Lukin3,Kuzmich} and of entanglement \cite{Choi}
or the first demonstration of functional network following the DLCZ
approach \cite{Kimble,Pan,Vuletic}. In the continuous variable
regime, which is considered here, storage of non-commuting quantum
variables of a light pulse has been demonstrated using the QND
scheme \cite{Polzik3}. Very recently, two experiments demonstrated
the storage of a squeezed light pulse with partial retrieval of the
squeezing \cite{Kozuma,Lvovsky}. At present, while advances have
been achieved in the direction of a quantum memory, it is
interesting to investigate a variety of
 systems allowing more flexibility. In this paper, we
study an EIT scheme allowing storage and retrieval of a single
sideband quantum field on an adjustable frequency range, using
atomic Zeeman coherence. The optimal response of the medium for
storage can thus be adapted to the frequency to be stored by
changing the magnetic field, while keeping the width of the EIT
window rather narrow. If symmetrical sidebands are stored in
separate atomic ensembles, this method should allow the storage of
various quantum signals.

We consider a large ensemble of three-level atoms in a $\Lambda$
level configuration interacting with two fields close to resonance
with the atomic transitions. The protocol relies on a weak field
carrying the quantum signal to be stored, and a strong, classical
control field that makes the medium transparent by way of EIT for
the signal field \cite{Harris}. Due to the EIT process, the group
velocity for the signal field is strongly reduced \cite{Hau1} and
the signal pulse can even be stopped \cite{Hau2Phillips}. When the
signal pulse is entirely inside the atomic medium, it can be shown
that the two quadratures of the field can be mapped into the
transverse components of the ground state angular momentum. Models
developed in Refs~\cite{Lukin1,Lukin2,Dantan2,Dantan3} predict a
high storage and retrieval efficiency. Moreover, taking into
account all the noise sources, including the atomic noise
generated by spontaneous emission and spin relaxation, a crucial
result of the theoretical model developed in Ref.~\cite{Dantan2}
is the absence of excess noise in the process of storage and
retrieval, provided the optical depth of the medium is large
enough.

\begin{figure}[h]
\centering
\includegraphics[width=7.8cm]{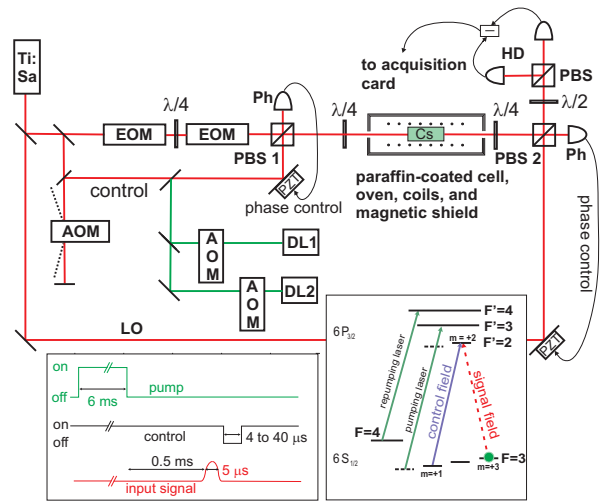}
\caption{Experimental set-up. DL 1 and 2 : pumping and repumping
diode lasers. HD : homodyne detection. Ph : photodiode. EOM :
electro-optical modulator. AOM : acousto-optical modulator. PBS :
polarizing beam splitter. The control field intensity is
controlled by an AOM, in double pass configuration in the
zero$^\mathrm{th}$ order (1:200 extinction). All laser beams have
a $1/e^{2}$ diameter of 14 mm in the cell. Insets : experimental
sequence and Cs transitions involved.} \label{setup}
\end{figure}
The experimental scheme is based on cesium vapor in a magnetic field
and uses the $6S_{1/2}$, F=3 to $6P_{3/2}$, F'=2 transition. The
control beam is $\sigma^{+}$ polarized and resonant with the
$m_{F}$=1 to $m_{F'}$=2 transition while the signal field is
$\sigma^{-}$ polarized and resonant with the $m_{F}$=3 to $m_{F'}$=2
transition (Fig.~\ref{setup} inset). In order to fulfill the
two-photon resonance condition, the detuning $\Omega$ between the
control and signal is equal to the Zeeman shift $2\Omega_{L}$
between the two considered sub-levels. By tuning the magnetic field,
one can optimize the memory response for a given frequency, allowing
a widely tunable frequency range for the signal to be stored.

The experimental set-up is shown in Fig.~\ref{setup}. The cesium
vapor is contained in a 3 cm long cell with a paraffin coating that
suppresses ground state relaxation caused by collisions with the
walls. The cell is heated to temperatures ranging from 30$^{o}$C to
40$^{o}$C, yielding optical depths from 6 to 18 on the signal
transition. It is placed in a longitudinal magnetic field produced
by symmetrical sets of coils and in a magnetic shield made of three
layers of $\mu$metal. Residual magnetic fields are smaller than
0.2~mG, with a homogeneity better than 1:700. The atoms are
optically pumped from the F=4 to the F=3 ground state and into the
$m_{F}=3$ sublevel of the F=3 ground state using diode lasers (2~mW
and 0.2~mW respectively).

The control beam is produced by a stabilized Titanium-Sapphire
(Ti:Sa) laser, with a linewidth of 100 kHz. The signal is a single
sideband field frequency shifted from the laser frequency by a set
of two electro-optical modulators. This sideband is a very weak
coherent field, with a power on the order of a fraction of nanowatt,
an adjustable frequency detuning $\Omega$ and a polarization
perpendicular to that of the carrier. The carrier is filtered out by
a polarizing beamsplitter (PBS1 in Fig.~\ref{setup}) and used to
lock the signal to control field relative phase. The light going out
of the cell is mixed with a local oscillator and analyzed using a
homodyne detection, after eliminating the control beam with a
polarizing beamsplitter (PBS2 in Fig.~\ref{setup}). The local
oscillator phase is locked to the one of the control beam after
PBS2. In the experimental sequence, shown in Fig.~\ref{setup}
(inset), the atoms are first optically pumped for 6 ms into the
$m_{F}$=3 sublevel, with a 92$\%$ efficiency. After a dark period of
0.5~ms, a 1.6 to 5~$\mu$s long signal pulse is sent into the cell
for the writing procedure. The control field is then switched off
for 4 to 40~$\mu$s, and finally turned on again.

\begin{figure}[h]
\centering
\includegraphics[width=7.8cm]{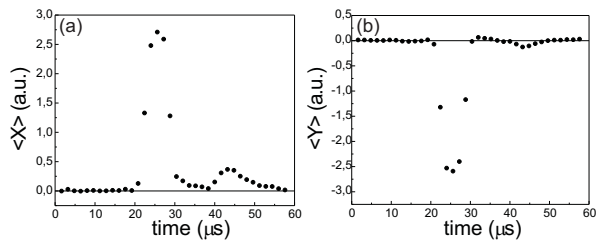}
\caption{Time-dependent mean values of the amplitude (a) and phase
(b) quadratures, measured from a typical 2000-sequence run. Cell
temperature T=40$^\mathrm{o}$C, control field power = 10~mW, input
pulse duration : 5~$\mu$s, input intensity : 0.1~nW.} \label{signal}
\end{figure}
The photocurrent difference from the homodyne detection is recorded
at a rate of 50~x~$10^{6}$ samples per second with a 14-bit
acquisition card (National Instruments NI 5122). A Fourier transform
is performed numerically by multiplying the signal with a sine or a
cosine function of frequency $\Omega$ and integrating over a time
$t_{m}=n 2 \pi /\Omega$, with $n$=2 to 4. This yields sets of
measured values of the quadrature operators $\hat{X}$ and $\hat{Y}$
of the outgoing field. Averaging over 2000 realizations of the
experiment gives the quantum mean values $<\hat{X}>$ and $<\hat{Y}>$
and variances $<(\Delta\hat{X})^2>=<(\hat{X})^{2}>-(<\hat{X}>)^{2}$
and $<(\Delta\hat{Y})^2>$ of the field quadratures. Typical traces
for mean values are shown in Fig.~\ref{signal}, for a storage time
of about 15~$\mu$s. The first peak corresponds to the leakage of the
signal field, the second one to the retrieved signal, for the
in-phase ($X$) quadrature (a) and for the out-of-phase ($Y$)
quadrature (b).

\begin{figure}[h]
\centering
\includegraphics[width=7.8cm]{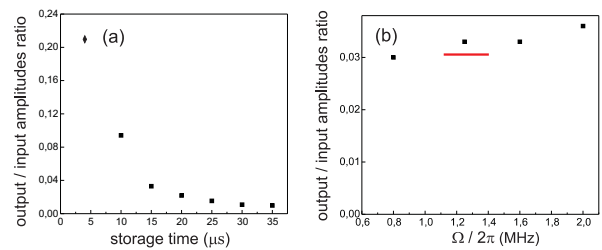}
\caption{(a) : Ratio of the amplitudes of the input and output
states, as a function of the storage time. T=40$^\mathrm{o}$C ;
The points indicated by ($\small{\blacksquare})$ ($\blacklozenge$)
correspond to a signal pulse duration of 6.4 $\mu \mathrm{s}$ (1.6
$\mu \mathrm{s}$) with a control field power of 10~mW (140 mW).
(b) : Ratio of the amplitudes of the input and output states, as a
function of the modulation frequency $\Omega / 2 \pi$. The bar
indicates the spectral width of the input pulse.} \label{st}
\end{figure}
Let us first discuss the results on the mean values. The
experiment clearly allows to store and retrieve the two
quadratures of a signal in the atomic ensemble. The storage
efficiency as a function of storage time is shown in Fig.~\ref{st}
(a). It decreases rapidly with the storage time, with a time
constant $\tau_{m} \sim 10\mu \mathrm{s}$, due to fast spin
decoherence processes in the ground state, which necessitates
further investigation. An efficiency of 21$\%$ has been measured
for a short storage time and a high value of the control field.

\begin{figure}[h]
\centering
\includegraphics[width=7.8cm]{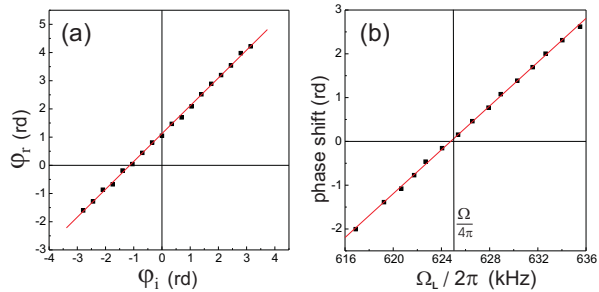}
\caption{Phase $\varphi_{r}$ of the retrieved pulse (a) as a
function of the input pulse phase $\varphi_{i}$ with a constant
$\Omega_{L}$, and (b) as a function of $\Omega_{L}$, with a fixed
input phase. The storage time is 20~$\mu$s, and $\Omega /
4\pi=$625~kHz}. \label{phase}
\end{figure}
In order to check the coherence of the process, we have performed a
detailed study of the phase of the retrieved signal
\cite{coherence}. Figure~\ref{phase}(a) shows the measured
dependence of the phase $\varphi_{r}$ of the retrieved pulse on the
phase $\varphi_{i}$ of the initial pulse. Phase $\varphi_{r}$ has a
linear dependence on $\varphi_{i}$ with a unit slope, confirming the
coherence of the process, while the non-zero value of $\varphi_{r}$
for $\varphi_{i}=0$ corresponds to a phase shift accumulated by
atomic coherence during the storage process. The latter is due to a
non-zero detuning $\delta=2\Omega_{L}-\Omega\neq0$ between the
atomic coherence and the two-photon transition, $\Omega_{L}$ being
the Larmor frequency. In Fig.~\ref{phase}(b) we show the measured
phase shift of the retrieved signal as a function of $\delta$ by
varying the magnetic field for a fixed storage time. The phase shift
has a linear dependence on the Larmor frequency, with a slope of
0.25 rd/kHz which is in very good agreement with the predicted
dependence, given by $\varphi_{r}=(2\Omega_{L}-\Omega)\tau $, where
$\tau$= 20~$\mu $s is the storage time.

\begin{figure}[h]
\centering
\includegraphics[width=8cm]{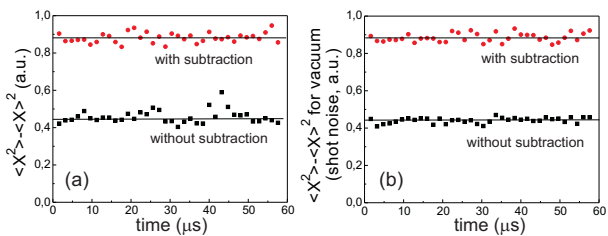}
\caption{(a) (resp. b) : variance of the amplitude quadrature for
the signal (resp. for vacuum). The phase quadrature variance is
similar.} \label{variances}
\end{figure}
The  noise curves corresponding the mean values discussed above
are obtained by calculating the variances from the same data set.
Because of a small leak of the control field into the signal field
channel, the raw data exhibit additional features due to the
transients of the control field. Although it has been designed
with a smooth shape, the control field contains Fourier components
around $\Omega/2\pi=1.25$ MHz. To get rid of this spurious effect,
the transients are measured independently after each sequence with
no signal field and the corresponding data are subtracted point to
point from the data taken with a signal field. The signal curves
shown in Fig.~\ref{signal} are obtained with this method. For the
noise, this procedure is equivalent to a 50/50 beamsplitter on the
analyzed beam and adds one unit of shot noise to the noise
measured without subtraction, yielding the upper curve in
Fig.~\ref{variances}(a). The noise calculated from the raw data,
without subtraction, corresponding to the lower curve in
Fig.~\ref{variances}(a), exhibits small additional fluctuations
when the control field is turned on for read-out. With the
subtraction procedure, these fluctuations are suppressed, showing
that they originate from a classical, reproducible spurious
effect.

The noise curves can be compared to the shot noise, which is
obtained independently from the same procedure with no control
field and no signal field in the input, without and with
subtraction, as shown in Fig.~\ref{variances}(b). The recorded
variances shown in Fig.~\ref{variances}a are found to be at the
same level as shot noise (Fig.~\ref{variances}b), which indicates
that the writing and reading processes add very little noise,
which can be evaluated to less than 2$\%$. Excess noise has been
studied by other authors \cite{Lam,Hetet} and it is a critical
feature for the storage benchmark. The noise curves of
Fig.~\ref{variances}a correspond to moderate values of the control
field power (10~mW). When the control field power is set to higher
values, excess noise appears. It originates from fluorescence and
coherent emission due the control beam \cite{Lvovsky} and from
spurious fluctuations from the turn on of the control field
leaking into the signal channel, that cannot be eliminated with
the subtraction procedure.

Following Ref.~\cite{Hetet}, we can evaluate the performance of our
storage device using criteria derived from the T-V characterization.
 This method had also been used to characterize quantum
non demolition measurements or teleportation~\cite{Roch}. The
conditional variance $V$ of the signal field quadratures before and
after storage is the geometrical mean value of the input-output
conditional variances $V=\sqrt{V_{X}V_{Y}}$ of the two quadratures,
with
$V_{X}=V_{X}^{out}-\frac{|\langle\hat{X}^{in}\hat{X}^{out}\rangle|^{2}}{V_{X}^{in}}$
and the same for $V_{Y}$, where $V_{X}^{in/out}$ is the variance of
the normalized input/output field quadratures denoted
$\hat{X}^{in/out}$. The transmission coefficient $T$ is the sum of
the transmission coefficients for the two quadratures
$T=T_{X}+T_{Y}$, with
$T_{X}=\mathcal{R}_{X}^{out}/\mathcal{R}_{X}^{in}$, where
$\mathcal{R}_{X}^{in/out}$ is the signal to noise ratio of
input/output field for the $X$ quadrature $\mathcal{R}_{X}^{in/out}=
\frac{4(\alpha_{X}^{in/out})^{2}}{V_{X}^{in/out}}$, with
$\alpha_{X}^{in/out}$ the coherent amplitude of the fields.

The experimental results shown in Fig.~\ref{signal} and
\ref{variances}, with a storage amplitude efficiency of 10$\%$,
corresponds to $T=0.02$ and $V=0.99$, if one considers that the
noise is equal to shot noise, and $V=1.01$ if the noise is 2$\%$
higher than shot noise. A classical memory with the same $T$
yields $V=1.01$. So, in this case, the performances of our system
are within the limit of the quantum domain. When the storage
amplitude efficiency is 21$\%$, which is obtained with a smaller
storage time and a higher control field, $T=0.08$, while the value
of the conditional variance of a system with such a $T$ without
excess noise is $V=0.96$ and the conditional variance for a
classical memory is $V=1.04$. The excess noise in our experiment
in that case is over 10$\%$, too large to reach the quantum
domain.

\begin{figure}[h]
\centering
\includegraphics[width=7.8cm]{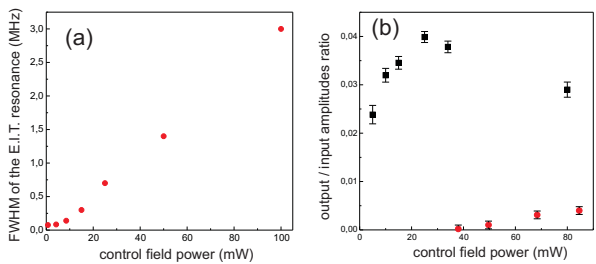}
\caption{ (a) FWHM of the EIT transparency window (MHz) as a
function of the control field power (mW). The control and signal
beams waist is $7.2$~mm. (b) ratio of the output and input states
amplitudes, as a function of the control field power, for a single
sideband of frequency 1.25 MHz (squares) and a dual sideband
modulation of frequency 400 kHz (circles). T= 50~$^\mathrm{o}$C .
Pulse duration : 5 $\mu \mathrm{s}$. Storage time : 15~$\mu
\mathrm{s}$.} \label{width_power}
\end{figure}

The method presented here concentrates on the storage of a single
sideband, which allows more flexibility than storing two symmetrical
sidebands in the same EIT window, especially for high frequency
components. We have measured the efficiency of the process when the
sideband frequency $\Omega$ is varied, as can be seen on
Fig.~\ref{st} (b). No significant variation with $\Omega$ is
observed. This comes from the fact that the frequency of the signal
to be stored can be matched to the position of the EIT window,
without changing its width. The EIT window width can be kept below 1
MHZ, as shown in Fig.~\ref{width_power}(a).

While many properties of quantum fields, such as squeezing, involve
two symmetrical frequency sidebands, manipulation of individual
sidebands provides an interesting tool for quantum information
processing. The photocurrent difference after the homodyne
detection, as measured here, yields the amplitude modulation
operator $\hat{X}(\Omega)$,
\begin{equation}
\hat{i}=\hat{X}(\Omega)=(\hat{X}_{\Omega}+\hat{X}_{-\Omega})\cos\Omega
t+ (\hat{Y}_{\Omega}-\hat{Y}_{-\Omega})\sin\Omega t \label{X}
\end{equation}
which is expressed as a combination of the quadrature operators of
the two sidebands, $\hat{X}_{\pm\Omega}$ and
$\hat{Y}_{\pm\Omega}$. In our present case, with a single
sideband, the observables to be measured, $\hat{X}_{\Omega}$ and
$\hat{Y}_{\Omega}$, are mixed with an empty sideband at $-\Omega$,
adding one unit of shot noise, which is intrinsic to homodyne
detection. In the case of a squeezed field, we have
$\Delta^{2}\hat{X}(\Omega)<1$ and Eq.~\ref{X} shows that the two
sidebands are entangled, as known for a long time \cite{Caves}.

To store a squeezed field at a given frequency $\Omega$, one has to
store its two entangled sidebands at $+\Omega$ and $-\Omega$, which
can be achieved by extending our method, first separating the two
sidebands using a Mach-Zehnder interferometer \cite{Leuchs} and
storing them into two atomic ensembles. The procedure results in
entangling the two ensembles. After read-out, the two sidebands can
eventually be recombined and measured with no vacuum added.

The two sidebands of a field can also be stored at the same time
in one atomic ensemble if $2\Omega$ is smaller than the width of
the EIT transparency window. Figure~\ref{width_power}(a) shows the
EIT linewidth in our case as a function of the control field
power. We have measured the efficiency of the storage process for
a signal field made of two sidebands at $\pm$400~kHz. The result
is shown in Fig.~\ref{width_power}(b). The efficiency is lower
than in the single sideband case, which can be attributed to the
large value of the time-bandwith product of the pulse to be
stored.

In conclusion, we have demonstrated storage and retrieval of the
two non-commuting quadratures of a small coherent state in an
atomic medium, with an excess noise small enough to put it inside
the quantum domain. This coherent state, made of a single sideband
of the control field, has been stored in the Zeeman coherence of
the atoms. Tuning the magnetic field allows to adjust the response
of the medium to the signal frequency, keeping the EIT window
rather narrow. Comparison with the storage of a modulation made of
two symmetrical sidebands shows the latter is hampered by the
finite bandwidth of the EIT window. Storing the two sidebands in
two separate atomic ensembles is thus a promising method for
quantum memory with a widely adjustable frequency. The method also
opens the way to the storage of multiplexed quantum signals, a
promising direction to increase the capacity of a quantum network.

The authors thank Xiao Jun Jia for valuable contributions. This
work was supported by the E. U. grants COVAQIAL and COMPAS, by the
French ANR contract IRCOQ and by the Ile-de-France programme
IFRAF.

$^\ast$ email : elg$@$spectro.jussieu.fr

\end{document}